\documentclass{npcs}
\usepackage{graphicx,amssymb}

\newcommand{\nuN}{$\nu N$}
\newcommand{\SF}{\emph{SF}}
\newcommand{\SFs}{\emph{SFs}}
\newcommand{\DIS}{\emph{DIS}}
\newcommand{\DL}{\emph{DL}}
\newcommand{\QCD}{\emph{QCD}}
\newcommand{\pQCD}{\emph{pQCD}}
\newcommand{\HENA}{\emph{HENA}}
\newcommand{\DLC}{\emph{DL+CTEQ5}}
\newcommand{\LGC}{\emph{Log+CTEQ5}}

\begin{document}

\runauthor{A. Z. Gazizov, S. I. Yanush} \runtitle{Small-x growth
of nucleon structure functions and its manifestations in
ultrahigh-energy neutrino astrophysics}

\begin{topmatter}

\title{Small-$x$ Growth of Nucleon Structure Functions and Its
Manifestations in Ultrahigh-Energy Neutrino Astrophysics}

\author{A. Z. Gazizov and S. I. Yanush}
\institution{B. I. Stepanov Institute of Physics of the National
Academy of Sciences of Belarus}
\address{F. Skariny Ave.\ 68, 220072 Minsk, Belarus}
\email{gazizov@dragon.bas-net.by}
\email{yanush@dragon.bas-net.by}

\begin{abstract}
Rapid growth of neutrino-nucleon cross-sections at high energies
due to hypothetical hard pomeron enhancement of \nuN-structure
functions is discussed. Differential and integral hadron moments,
which, together with cross-sections, define rates of
hadron-electromagnetic cascades in a neutrino detector are
calculated for different power-law decreasing neutrino spectra.
For comparison two small-$x$ extrapolation schemes are discussed.
First includes Regge theory inspired hard pomeron enhancement of
\nuN-structure functions. The second is obtained with the help of
trivial extrapolation of perturbative \QCD\ structure functions
from the large $x$ region to the small $x$ one.

Implications of hard pomeron effects for cross-sections and hadron
moments are demonstrated. The most pronounced manifestations are
found in integral hadron moments for the case of charged current
electron (anti)neutrinos scattering off nucleons.
\end{abstract}

\end{topmatter}

\section{Introduction}
This paper continues the discussion of some specific features of
\nuN\ \emph{Deep Inelastic Scattering (DIS)} at extremely high, up
to $E_\nu \sim 1\times 10^{12}$~GeV, energies that was started at
the previous Seminar \emph{'NPCS-2000'} \cite{GYNPQS}. The main
idea of our approach is to extend the successful small-$x$
description of $F_2^{ep}(x,Q^2)$ \emph{Structure Function (SF)} by
A. Donnachie and P. V. Landshoff \emph{(DL)} \cite{DL} to
\nuN-\SFs, namely, to $F_2^{\nu N}(x,Q^2)$, $F_3^{\nu N}(x,Q^2)$.
\DL\ claim that record small-$x$ $ep$-scattering data by
\emph{HERA} \cite{HERA} may be successfully explained with the
help of a simple combination of several Regge theory inspired
non-perturbative pomerons. The most important are \emph{'soft'}
pomeron (with intercept $\sim 1.08$) and \emph{'hard'} pomeron
(with intercept $\approx 1.4$). First prevails at small $Q^2$,
while the latter dominates at large $Q^2$. Moreover, \DL\ argue
that perturbative \QCD\ (\pQCD) fails at small $x < 10^{-5}$ and
that its validity at $x \sim 10^{-4}\div 10^{-5}$ is a pure fluke.
However, one should keep in mind that \DL's approach neglects the
other, non-leading, poles and cuts in the complex angular momentum
$l$-plane. This common feature of pomeron physics causes this
model to violate the unitarity at $E_\nu \rightarrow \infty$.

Using a nontrivial generalization of \DL's $F_2^{ep}(x,Q^2)$ \SF\
description to the \nuN-scattering case, we have constructed
$F_2^{\nu N}(x,Q^2)$ and $F_3^{\nu N}(x,Q^2)$  \SFs, presumably
valid in the whole range of kinematic variables $0 \leq x \leq 1$
and $0 \leq Q^2 \leq \infty$ \cite{GYNPQS,GY}. At $x \gtrsim
10^{-5}$ they are chosen to coincide with \pQCD\ parameterization
by \emph{CTEQ5} collaboration \cite{CTEQ}, while in the small-$x$
region these \SFs\ are driven by the analogous Regge theory
inspired description. A special interpolation procedure, developed
in Ref.~\cite{GY}, allows to meet smoothly these different, both
over $x$ and $Q^2$, descriptions of \SFs\ at low and high $x$. In
Ref.~\cite{GYNPQS} these \SFs\ were denoted by \emph{DL+CTEQ5},
indicating that they have their origin in the interpolation
between \DL\ and \pQCD\ descriptions.

In parallel there were considered \SFs\ obtained via simple
extrapolation of \pQCD\ \SFs\ from $x \ge 1\times 10^{-5}$ to the
small-$x$ region:
\begin{equation}
\label{log} F_i^{\nu N,Log+CTEQ5}(x<x_{min},Q^2) = F_i^{\nu
N,CTEQ5}(x_{min},Q^2)\left(\frac{x}{x_{min}}\right)^{\beta_i(Q^2)}, %
\end{equation}
\begin{equation}
\label{beta}
\beta_i(Q^2) = \left. \frac{\partial \ln F_i^{\nu N, CTEQ5}(x,Q^2)}%
{\partial \ln x} \right|_{x=x_{min}} ; \;\;\; %
x_{min} = 1\times 10^{-5}.
\end{equation}%
These \SF's\ smoothly shoot to the low-$x$ region from the
\emph{CTEQ5} defined high-$x$ one \cite{BGZR,GYNPQS,GY}. Starting
values of functions and of their logarithm derivatives over $\ln
x$ are taken here at the $x=x_{min}$ boundary of \emph{CTEQ5}.
This parameterization was designated as \LGC.

Below we shall evaluate several observables involved in High
Energy Neutrino Astrophysics (\HENA) using both parameterization.
We shall compare them so that to reveal the manifestations of hard
pomeron enhancement in these values.

\section{Observables of HENA}
An incident cosmic high-energy $\nu$-flux can be detected only by
registration of secondary particles, the products of \nuN- and/or
$\nu e$-collisions with matter (basic ideas of \HENA\ are
expounded in Ref.~\cite{BBGGPfull}). Hence, the observables are
mostly rates of high-energy muons and/or of
nuclear-electromagnetic cascades. In this paper we shall discuss
only \nuN-interactions, though the most remarkable process in
\HENA\ is resonance cascade production via
\begin{equation}\label{nueres}
\bar{\nu}_e + e^- \rightarrow W^- \rightarrow \mbox{hadrons}
\end{equation}
at $E_{\bar{\nu}_e} \simeq 6.4 \times 10^{15}$~eV
\cite{ResJL,BGResSJNP}. This resonance should show itself as a
narrow high spike in the differential energy spectrum of cascades;
at resonance energy it exceeds essentially the ordinary
\nuN-interaction background.

Muons and cascades are produced via $CC$-
\begin{equation}\label{nuNscatCC}
  \nu_\mu (\bar{\nu}_\mu) + N \rightarrow \mu^\mp + X
\end{equation}
and $NC$-
\begin{equation}\label{nuNscatNC}
  \nu_\mu (\bar{\nu}_\mu) + N \rightarrow \nu_\mu (\bar{\nu}_\mu) + X
\end{equation}
\DIS. And in special case of electron (anti)neutrino
$CC$-interaction,
\begin{equation}\label{nueNscat}
  \nu_e (\bar{\nu}_e) + N \rightarrow e^{\mp} + X,
\end{equation}
both final states contribute to the same cascade, so that the
whole energy of an incident neutrino is transferred in it. But the
$NC$-scattering case of $\nu_e (\bar{\nu}_e) N$ does not differ
from (\ref{nuNscatNC}).

Differential and integral rates of cascade production in a
detector for a model neutrino flux with a power-law decreasing
energy spectrum,
\begin{equation}\label{spectra}
F_\nu(E_\nu) = A \times E_\nu^{-(\gamma+1)},
\end{equation}
where $\gamma$ is the index of an integral neutrino spectrum
$F_\nu(>E)$ (it is commonly assumed that $1.1 \leq \gamma \leq
2.1$), may be calculated with the help of so-called differential,
\begin{equation}\label{dhadrmom}
Z_h(E_h,\gamma) = \int \limits_0^1 dy y^\gamma \frac{d\sigma_{\nu
N}(E_h/y,y)}{\sigma_0 dy},
\end{equation}
and integral,
\begin{equation}\label{ihadrmom}
Y_h(E_h,\gamma) = \gamma \int \limits_0^1 du u^{\gamma-1}
Z_h\left(\frac{E_h}{u},\gamma\right),
\end{equation}
hadron moments \cite{NuNBG,Numom}. Here $E_h$ is the energy of a
hadron-electromagnetic cascade, $y = E_h/E_\nu$ and $\sigma_0$ is
the normalization cross-section; for $m_W = 81$~GeV $\sigma_0 =
1.09 \times 10^{-34}$~cm$^2$. These rates in $CC$-scattering case
(\ref{nuNscatCC}) are
\begin{equation}\label{drateofcasc}
  dN_h(E_h)/dt = Z_h^{CC}(E_h,\gamma) N_N \sigma_0 \Omega F_\nu(E_h),
\end{equation}
\begin{equation}\label{irateofcasc}
  dN_h(>E_h)/dt = Y_h^{CC}(E_h,\gamma) N_N \sigma_0 \Omega F_\nu(>E_h).
\end{equation}
Here $N_N$ is the number of nucleons in a detector and $\Omega$ is
an effective solid angle the neutrino flux comes from. The cases
of $\bar{\nu} N$-  and $NC$-interactions may be accounted for in a
similar way: one is to substitute in (\ref{dhadrmom}) the
appropriate differential cross-section for the $CC$ one.

Note that in Eq.s~(\ref{drateofcasc},\ref{irateofcasc}) both
differential and integral neutrino fluxes are taken at the cascade
energy $E_h$ and that for power-law decreasing spectra
(\ref{spectra}) the following useful relation is valid:
\begin{equation}\label{relat}
\gamma \times F_\nu(>E) = E \times F_\nu(E).
\end{equation}

In the case of (anti)neutrino $CC$-scattering (\ref{nueNscat}) a
role of differential hadron moment belongs to the normalized
$CC$-cross-section, $\sigma_{\nu N}^{CC}(E_\nu)/\sigma_0$; since
now $E_h=E_\nu$, $\nu_e$-flux in
(\ref{drateofcasc},\ref{irateofcasc}) is to be taken at the energy
of incident neutrino.

\section{Cross-sections of \nuN-scattering}
\begin{figure}[h!]
\begin{minipage}[t]{.475\linewidth}
\includegraphics[width=\linewidth]{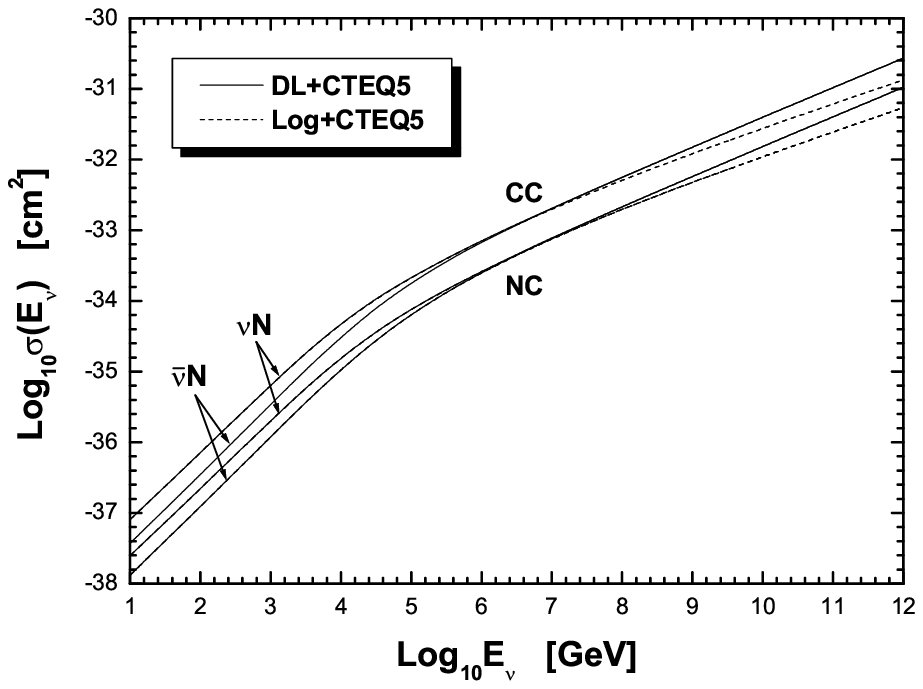}
\caption{\nuN\ and $\bar{\nu} N$ cross-sections calculated within
\emph{DL+CTEQ5} and \LGC\ models for $CC$- and $NC$-interactions.}
\label{fig:crsscn}
\end{minipage}\hfill
\begin{minipage}[t]{.475\linewidth}
\includegraphics[width=\linewidth]{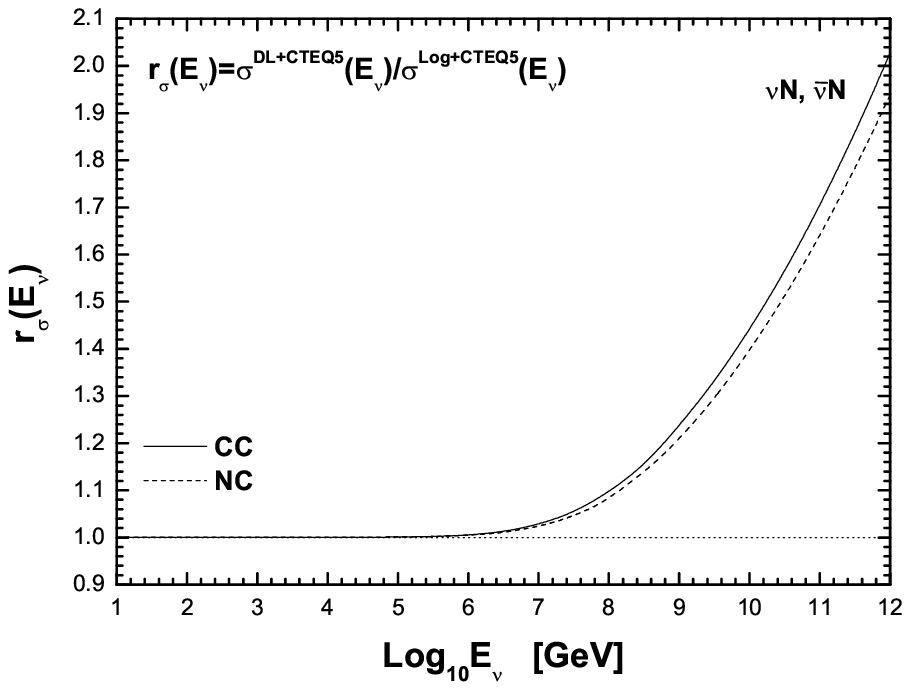}
\caption{Ratios of corresponding \emph{DL\-+CTEQ5} and \LGC\
cross-sections for \nuN- and (indistinguishable here) $\bar{\nu}
N$-scattering cases.} \label{fig:rsigma}
\end{minipage}
\end{figure}

In parton picture \nuN-cross-section increases with the energy due
to multiplication in number of the nucleon small-$x$ 'sea'-quark
contents. In the framework of non-perturbative pomeron approach
such growth occurs due to specific poles in the complex $l$-plane.
Calculated within \DLC\ and \LGC\ parameterizations, $CC$ and $NC$
$\nu (\bar{\nu}) N$-cross-sections are shown in
Fig.~\ref{fig:crsscn}.
The most part of high-energy cross-section is accumulated at small
$x$ and high $Q^2$, where hard pomeron term dominates \SFs. Since
hard pomeron enhanced \DLC\ small-$x$ \SFs\ are higher than
corresponding \LGC\ ones, their cross-sections prevail over
'perturbative' at high energies. Discrepancies become especially
clear in Fig.~\ref{fig:rsigma}, where ratios of corresponding
\DLC\ and \LGC\ cross-sections, $r_\sigma(E_\nu)$, are plotted
versus neutrino energy. The curves for \nuN\ and $\bar{\nu} N$
cases practically coincide in this graph; ratios for
$NC$-interactions are very close to corresponding $CC$ ones.

It should be noted, that hard pomeron enhanced growth of
\nuN-cross-sections is the most rapid among all presently known,
which have been obtained under different 'ordinary' (no extra
dimensions and so on) assumptions (see Ref.~\cite{GY} and
references therein).

\section{Differential and integral hadron moments}
According to Eq.s~(\ref{dhadrmom},\ref{ihadrmom}), hadron moments
\begin{figure}[h!]
\begin{minipage}[t]{.475\linewidth}
\includegraphics[width=\linewidth]{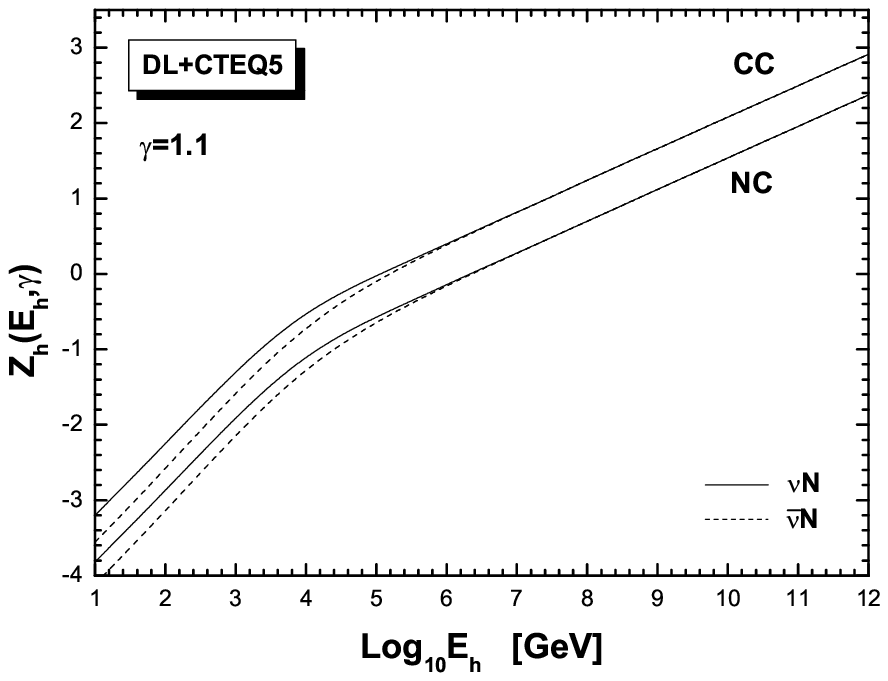}
\caption{\nuN\ and $\bar{\nu} N$ \DLC\ $CC$ and $NC$ differential
hadron moments for integral $\nu$-spectrum index $\gamma=1.1$.}
\label{fig:zh11}
\end{minipage}\hfill
\begin{minipage}[t]{.475\linewidth}
\includegraphics[width=\linewidth]{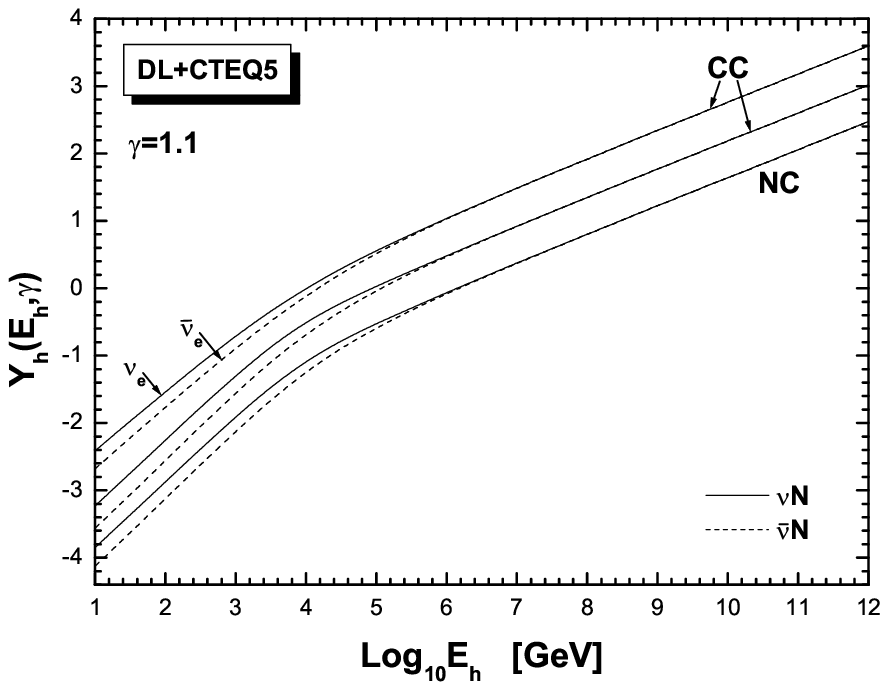}
\caption{The same curves as in Fig.~\ref{fig:zh11}, but for
integral hadron moments. $\nu_e N$- and $\bar{\nu}_e
N$}-scattering cases are shown separately. \label{fig:yh11}
\end{minipage}
\end{figure}
depend on high-energy part of $\nu$-spectrum where cross-sections
are higher. As a consequence, hard pomeron effects look even more
pronounced in these observables. To illustrate a common trend, the
differential and integral \nuN\ and $\bar{\nu} N$ hadron moments
are plotted in Fig.~\ref{fig:zh11} and Fig.~\ref{fig:yh11},
respectively, for \DLC\ parameterization and $\gamma = 1.1$, both
for $CC$- and $NC$-interactions.

\begin{figure}[h!]
\begin{minipage}[t]{.475\linewidth}
\includegraphics[width=\linewidth]{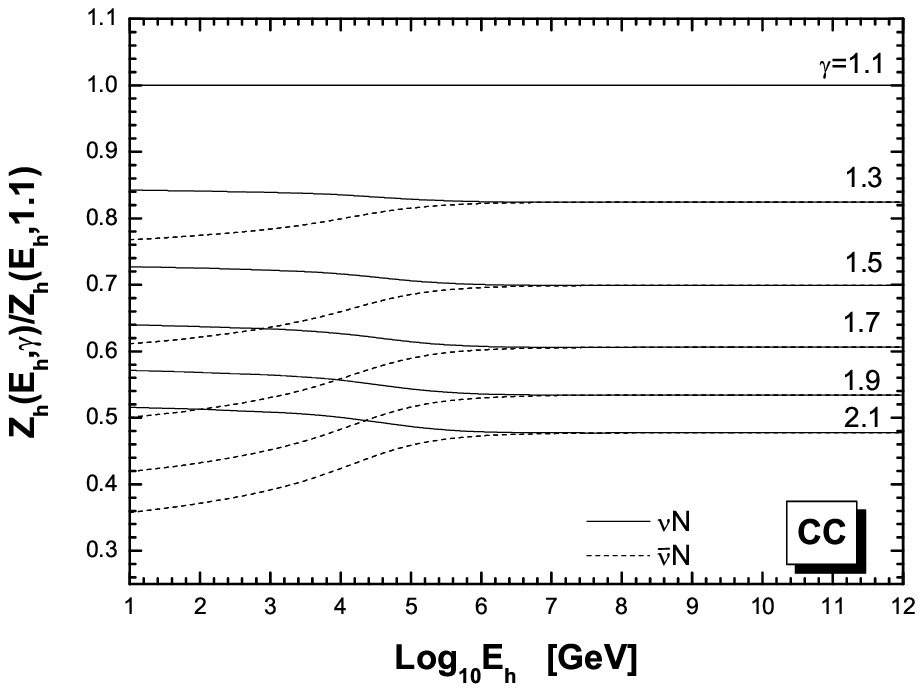}
\caption{Ratios for \DLC\ $CC$ differential hadron moments.}
\label{fig:rzcc}
\end{minipage}\hfill
\begin{minipage}[t]{.475\linewidth}
\includegraphics[width=\linewidth]{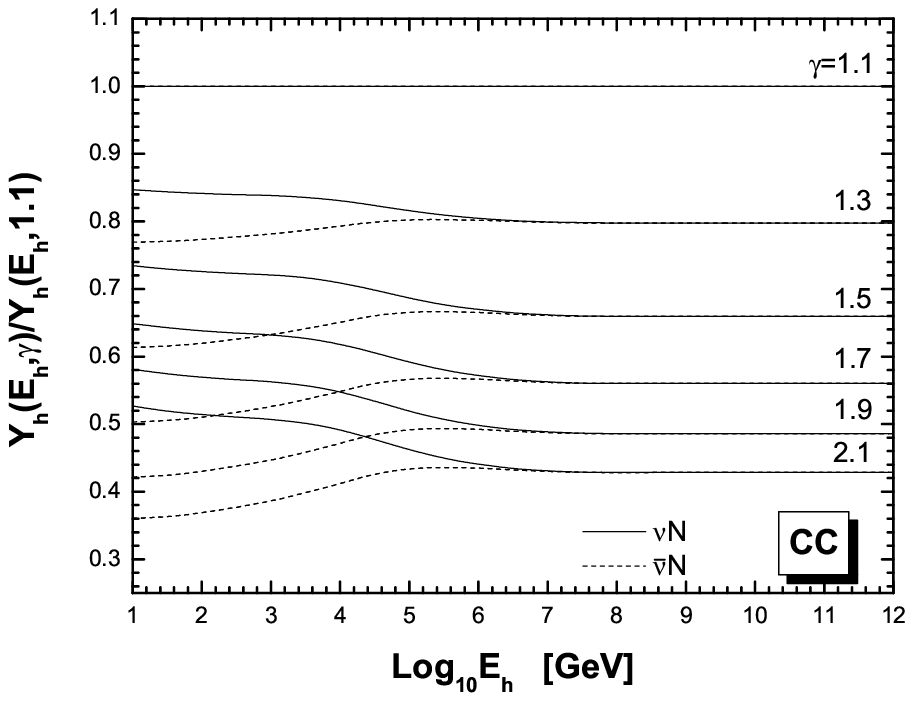}
\caption{Ratios for \DLC\ $CC$ integral hadron moments.}
\label{fig:rycc}
\end{minipage}
\end{figure}

\begin{figure}[h!]
\begin{minipage}[t]{.475\linewidth}
\includegraphics[width=\linewidth]{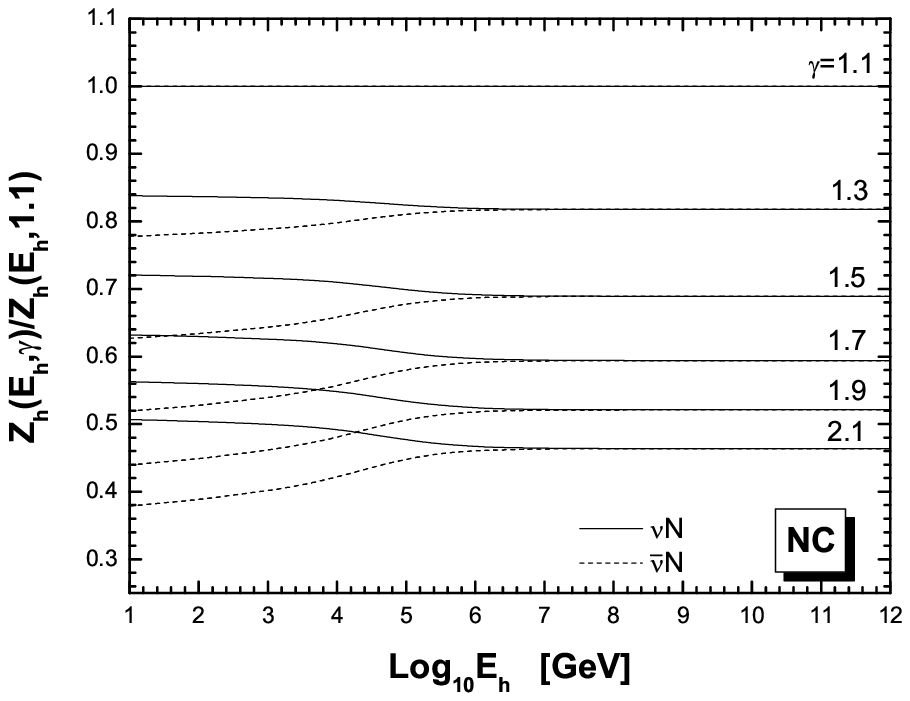}
\caption{Ratios for \DLC\ $NC$ differential hadron moments.}
\label{fig:rznc}
\end{minipage}\hfill
\begin{minipage}[t]{.475\linewidth}
\includegraphics[width=\linewidth]{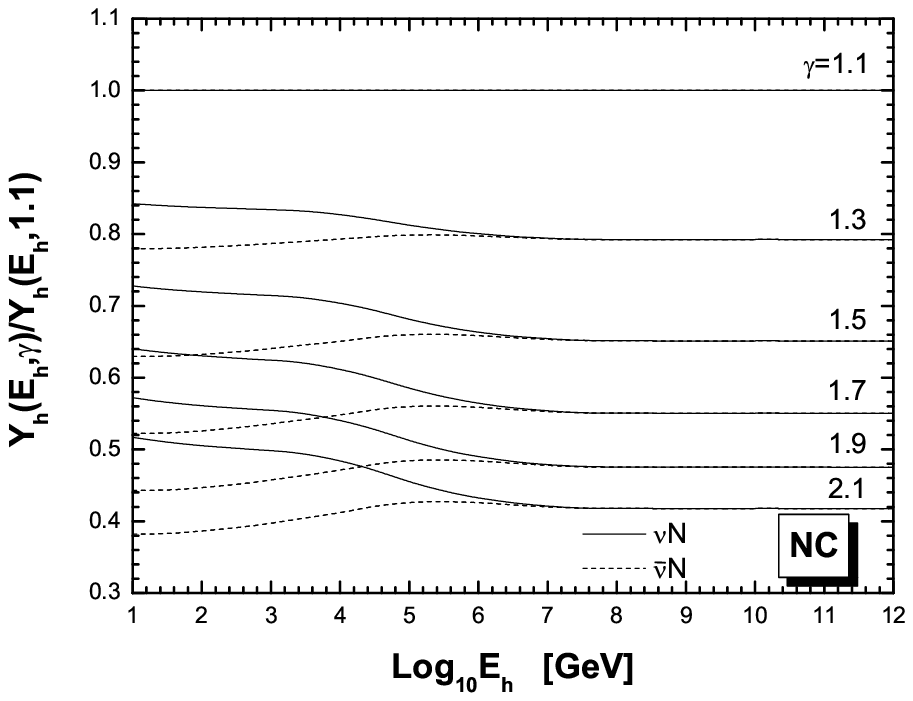}
\caption{Ratios for \DLC\ $NC$ integral hadron moments.}
\label{fig:rync}
\end{minipage}
\end{figure}

Dependencies of $CC$ and $NC$ hadron moments on $\gamma$ are shown
in
Fig.s~\ref{fig:rzcc},\ref{fig:rycc},\ref{fig:rznc},\ref{fig:rync}
with the help of ratios between corresponding moments with
indicated $\gamma$ and those with $\gamma=1.1$.

\begin{figure}[h!]
\begin{minipage}[t]{.475\linewidth}
\includegraphics[width=\linewidth]{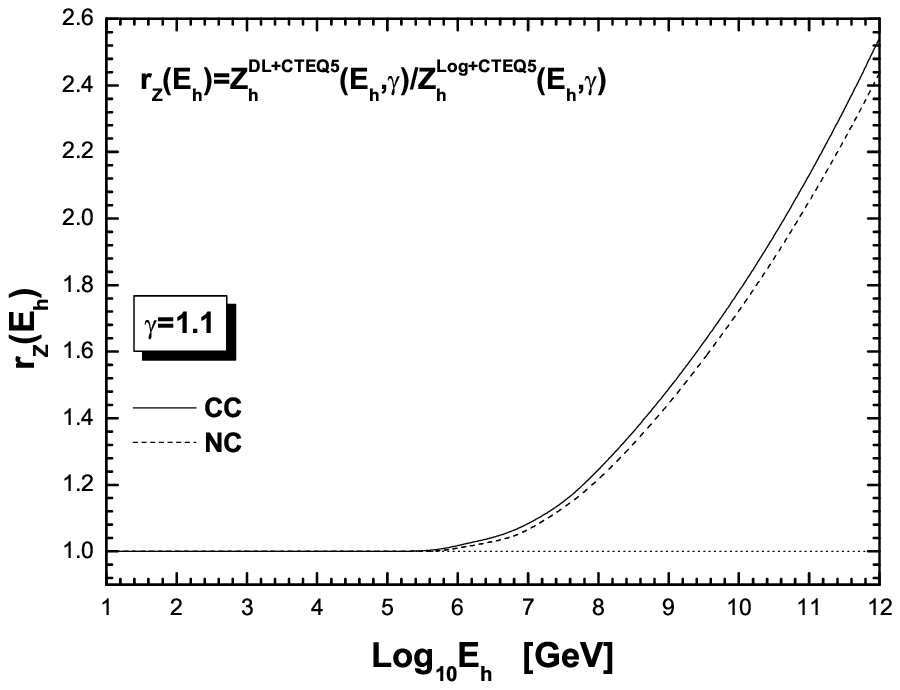}
\caption{Ratios between $CC$ differential hadron moments of \DLC\
and \LGC\ parameterizations for $\gamma=1.1$. } \label{fig:rzDL}
\end{minipage}\hfill
\begin{minipage}[t]{.475\linewidth}
\includegraphics[width=\linewidth]{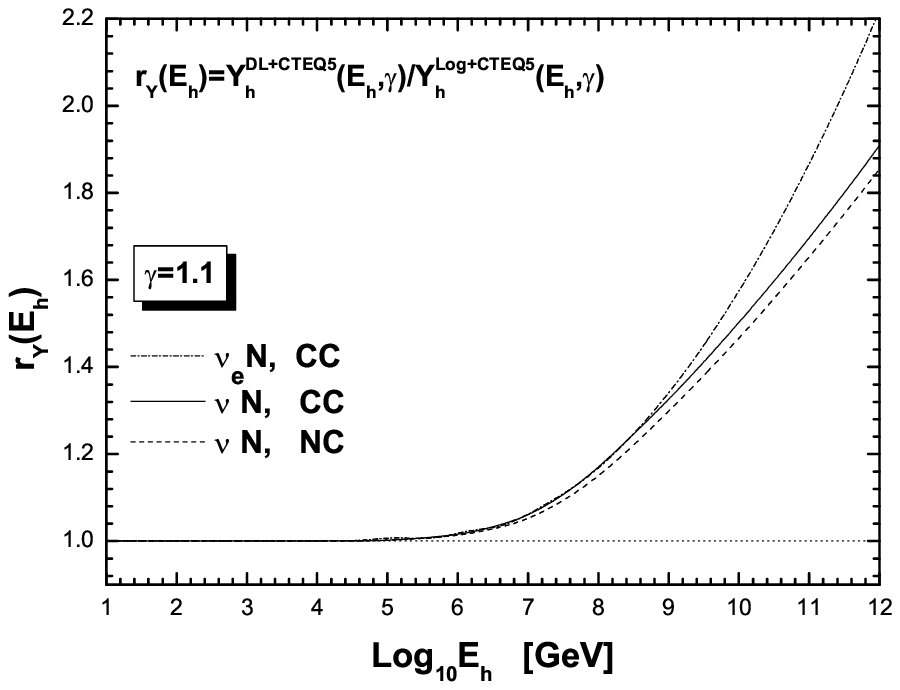}
\caption{The same as in Fig.~9, but for integral hadron moments.}
\label{fig:ryDL}
\end{minipage}
\end{figure}

And, finally, ratios between corresponding $CC$ hadron moments of
\DLC\ and \LGC\ parameterizations are demonstrated in
Fig.s~\ref{fig:rzDL},\ref{fig:ryDL} for $\gamma=1.1$. Due to
sensitivity to higher energies, these ratios are higher than
ratios of cross-sections. The most salient difference is seen in
$Y_h^{CC}(E_h,\gamma)$ for $\nu_e (\bar{\nu}_e) N$ scattering,
where the whole energy goes to the cascade.
\section{Conclusions}
We have demonstrated that small-$x$ hard pomeron enhancement of
\nuN\ structure functions evinces itself via essential growth of
some \HENA\ observables at high energies. For example,
cross-sections and hadron moments, defining rates of cascades,
grow more rapidly with the energy than in the case of trivial
\pQCD\ extrapolation. The calculated hadron moments may be used
for estimation of cascade rates in future giant high-energy
neutrino detectors.
\section*{Acknowledgments}
The authors are grateful to Prof.\ V. S. Berezinsky for his
partial participation, useful comments and encouragements. This
work was supported by the \emph{INTAS} grant No: 99-1065.

\end{document}